\documentclass[sigconf]{acmart}

\usepackage{booktabs} 
\usepackage{array}

\renewcommand\footnotetextcopyrightpermission[1]{} 
\begin{document}
\title[Linking user preference and news consumption]{Intelligent social bots uncover the link between user preference and diversity of news consumption}

\author{Yong Min}
\authornote{Corresponding author}
\orcid{}
\affiliation{
  \institution{Zhejiang University of Technology}
  \streetaddress{}
  \city{Hangzhou}
  \state{Zhejiang}
  \postcode{}
}
\email{myong@zjut.edu.cn}

\author{Tingjun Jiang}
\affiliation{
  \institution{Zhejiang University of Technology}
  \streetaddress{}
  \city{Hangzhou}
  \state{Zhejiang}
  \postcode{}
}
\email{jtj9678@163.com}

\author{Xiaogang Jin}
\orcid{}
\affiliation{
  \institution{Zhejiang University}
  \streetaddress{}
  \city{Hangzhou}
  \state{Zhejiang}
  \postcode{}
}
\email{xiaogangj@zju.edu.cn}

\author{Cheng Jin}
\orcid{}
\affiliation{
  \institution{Tencent Corporation}
  \streetaddress{}
  \city{Shenzhen}
  \state{Guangdong}
  \postcode{}
}
\email{moonjin@tencent.com}

\author{Qu Li}
\orcid{}
\affiliation{
  \institution{Zhejiang University of Technology}
  \streetaddress{}
  \city{Hangzhou}
  \state{Zhejiang}
  \postcode{}
}
\email{liqu@zjut.edu.cn}

\renewcommand{\shortauthors}{Y. Min et al.}

\begin{abstract}
The boom of online social media and microblogging platforms has rapidly alter the way we consume news and exchange opinions. Even though considerable efforts try to recommend various contents to users, loss of information diversity and the polarization of interest groups are still an enormous challenge for industry and academia. Here, we take advantage of benign social bots to design a controlled experiment on Weibo (the largest microblogging platform in China). These software bots can exhibit human-like behavior (e.g., preferring particular content) and simulate the formation of personal social networks and news consumption under two well-accepted sociological hypotheses (i.e., homophily and triadic closure). We deployed 68 bots to Weibo, and each bot ran for at least 2 months and followed 100 to 120 accounts. In total, we observed 5,318 users and recorded about 630,000 messages exposed to these bots. Our results show, even with the same selection behaviors, bots preferring entertainment content are more likely to form polarized communities with their peers, in which about 80\% of the information they consume is of the same type, which is a significant difference for bots preferring sci-tech content. The result suggests that users\textquotesingle preference played a more crucial role in limiting themselves access to diverse content by compared with the two well-known drivers (self-selection and pre-selection). Furthermore, our results reveal an ingenious connection between specific content and its propagating sub-structures in the same social network. In the Weibo network, entertainment news favors a unidirectional star-like sub-structure, while sci-tech news spreads on a bidirectional clustering sub-structure. This connection can amplify the diversity effect of user preference. The discovery may have important implications for diffusion dynamics study and recommendation system design.
\end{abstract}

%
%


\maketitle

\vspace{10pt}
``\textit{...forms of media favor particular kinds of content and therefore are capable of taking command of a culture.}''

\rightline{Neil Postman \cite{postman06}}

\section{Introduction}
With the growing popularity of social media and microblogging platforms, the way people obtain information and form opinions has undergone substantial change. A recent study found that social media has become the primary source for over 60\% of users to obtain news \cite{newman15}. These users are exposed to more personalized information, which is considered to limit the diversity of content they consume and even cause filter bubbles \cite{pariser11,flaxman16,zuiderveen16}. In fact, news consumption on social media has been extensively studied to determine what factors lead to polarization \cite{dandekar13,schmidt17}. Recent works suggest that confirmation bias or selective exposure plays a significant role in online social dynamics \cite{dandekar13,bakshy15}. That is, online users tend to select messages or information sources supporting their existing beliefs or cohering with their preferences and hence to form filter bubbles \cite{zuiderveen16}. However, the mechanism of polarization remains an open question that needs further clarification.

\begin{figure}
\includegraphics{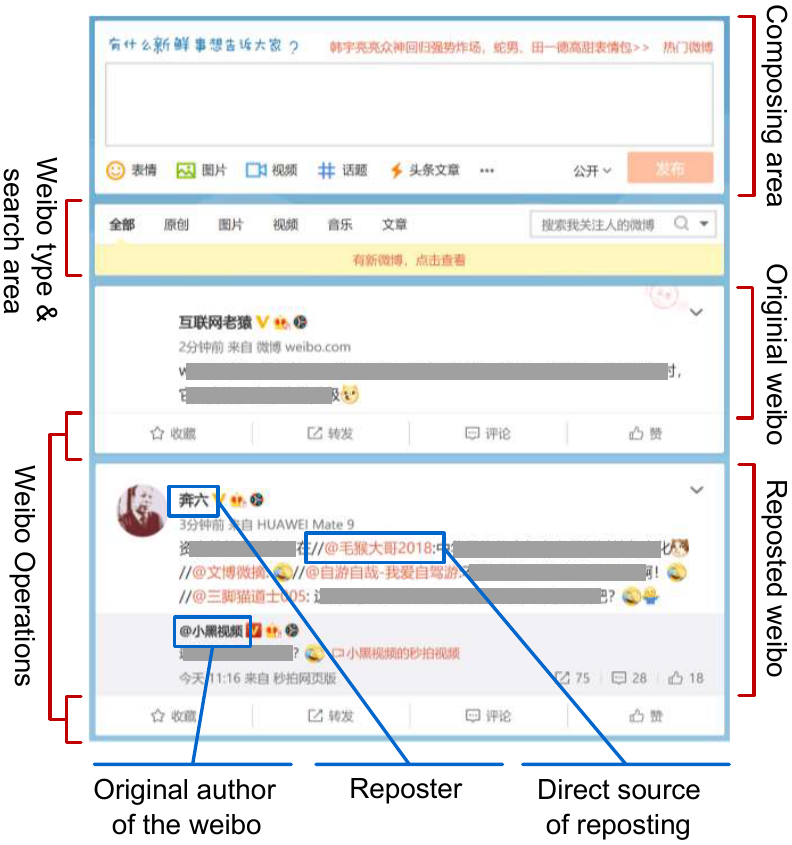}
\caption{Weibo user interface.}
\label{fig:weibo}
\end{figure}

In this paper, we conduct a controlled experiment to explore the pattern of news consumption and the factors leading to the emergence of polarization on Weibo (NASDAQ: WB), the most popular Chinese microblogging service. Weibo is the most famous Chinese social networking and microblogging platform where users can post messages for instant sharing. It has more than 500 million registered users and 313 million active monthly users, including government agencies, celebrities, enterprises, and ordinary netizens. The information dissemination across Weibo is chiefly based on the ``attention'' relationship between users \cite{wiki18}. User A can follow user B (i.e., becoming a ``follower'' of B, and B then becomes a ``following'' of A) and then receive all posts by user B on his page. Therefore, the social network of Weibo is a directed graph. Reposting or forwarding followings' messages are the primary methods of diffusing information in Weibo. One microblog can be reposted multiple times. In a reposted microblog, the usernames of the original author and the last re-poster are visible to the current user (Figure~\ref{fig:weibo}). We focus on two types of user preference (entertainment and sci-tech news) and analyze their effects on news consumption behaviors as well as the local network structure.

In order to explore the extent and pathway of preferences affecting news consumption, we propose an experimental approach by using intelligent social bots \cite{ferrara16}. Social bots are generally considered to be harmful, although some of them are benign and, in principle, innocuous or even helpful. Therefore, social bots are often the subject of research that needs to be eliminated \cite{paradise14}, but researchers have yet to recognize their potential value as a powerful tool in social network analysis \cite{abokhodair15,ferrara16}. Our social bots can use text classification algorithms to simulate the selection behavior of users on content and information sources, thereby simulating the evolution of personal social networks of users with specific preferences. By analyzing the personal social networks of these robots and the information obtained from them, we can clearly show the route of user preferences affecting information polarization, thus revealing the complex interactions behind information polarization.

\section{RELATED WORK}

\subsection{Polarization of news consumption}

The advent and growing popularity of social media and personalization services have, to a large extent, changed the way people consume information, which is regarded as a factor in the fostering of the polarization of information diffusion and public opinions \cite{bakshy15}. In collective and individual levels, the term ``polarization'' has two possible meanings. One is that like-minded people form exclusive clubs sharing similar opinions and ignoring dissenting views, i.e., opinion polarization \cite{dandekar13}; the other is that an individual is exposed to less diverse content and is limited by a narrow set of information, i.e., information polarization \cite{schmidt17}. For this paper, we focus on information polarization on social networks and media.

Many scholars have investigated the polarization of political news consumption \cite{schmidt17,conover11,markus13,narayanan18}. For example, Schmidt et al. (2017) found that users prefer to pay attention to a narrow set of news outlets \cite{schmidt17}. However, the phenomenon of information polarization is present not only in political content but also in other content, especially entertainment \cite{webster05}. When the overwhelming entertainment news obscures the scientific, financial and social news that reaches their home page daily, users are likely to become addicted to the ``entertainment gossip'' and consume more gossip-related news \cite{kim12}. A high proportion of these users are so-called fanboys or fangirls who desperately support their favorite celebrities, and some others are anti-fans of those celebrities \cite{duffett13}. Fans and anti-fans have similar features and behavior in that they are usually active in polarized clusters of like-minded people, maintaining and reinforcing their current standpoints and even attacking each other \cite{bennett13}. These polarized users could be the important force in making and spreading ridiculous rumors about celebrities because they want to look only for news that supports their views and does not care about the facts. Besides, most social networks are usually flooded with malicious attacks and profanities between fans and anti-fans, making the network environment increasingly worse \cite{huang16}. Moreover, these extreme behaviors gradually extend into real life. For instance, some netizens unearth the real identity of people who hold opposing views by launching a search and then exposing personal information (including photos, cell phone number, home address, and other details) to disturb their opponent's real life \cite{han18}. These phenomena reveal the potential interaction between information polarization in online social networks and offline behavioral polarization.

The formation mechanism of polarization and the method of restraining polarization are the most pressing issues in the field of information polarization \cite{stroud10,pariser11,bakshy15,zuiderveen16}. Many studies have suggested that personalization, both self-selection, and pre-selection, limit people\textquotesingle s exposure to diverse content and increases polarization \cite{kakiuchi18}. Self-selection is the tendency for users to consume content or build new relationships that confirm their existing beliefs and preferences. Pre-selection depends on computer algorithms to personalize content for users without any conscious user choice. The present research generally agreed that self-selection is the primary cause of information polarization \cite{bakshy15}.

\subsection{Controlled experiment on online social networks}

At present, research about online social networks relies mainly on observational methods \cite{bakshy15,schmidt17}, which means that researchers cannot intervene in the study object but can only process and analyze the passively acquired data. However, the big open data contains noise that can affect the results \cite{aral14}. To overcome the shortcomings of a purely observational study, some researchers have ingeniously adopted the method of natural experiments or quasi-experiments to make comparative analyses and causal inferences from the available datasets \cite{chen11,phan15,vosoughi18}. For example, Phan et al. (2015) analyzed the dynamic development of social relationships after natural disasters in the context of Hurricane Ike in 2008 \cite{phan15}. Nevertheless, the uncontrolled approach limits the freedom of research. In contrast, the controlled experiment is more rigorous and can provide a more solid foundation for causal analysis \cite{aral14}.

As online social media and networks have expanded to serve hundreds of millions of users, and their functions have become more accessible and intelligent, the ability to test sociological theories and behavioral phenomena has increased. Application programming interfaces from operators and innovative third-party toolkits allow researchers to execute novel experimental designs. For example, Facebook provides interfaces to customize application features for particular users and enables the design of controlled experiments \cite{aral12}. Amazon Mechanical Turk lets experimenters engaging worldwide users in a precisely defined experimental environment \cite{rand11b,mason12}. These powerful tools are enabling a revolution of experimental social science and social network analysis. As a result, the controlled experiment approach has begun to reveal the nuanced causal relationships among the design and operation of the social network and human behavior. For example, recent large-scale experiments have deepened our understanding about information sharing and diffusion, behavior spreading, voting and political mobilization, and cooperation \cite{centola10,centola11,rand11a,bond12,kramer14,aral14}. The large scale of modern online social networks also enables researchers to conduct experiments involved millions of people to explore the variety of behaviors of diverse users. For example, the experiment conducted by Aral and Walker (2012) estimates the heterogeneity of the impact of influence-mediating messages on different types of consumers \cite{aral12}; Muchnik et al. (2013) examine in depth whether opinion change or selective turnout creates the social influence bias that they estimate exists in online ratings \cite{muchnik13}.

Large-scale experiments on social networks, while actively promoting innovation of sociology and communication studies, also bring about some serious problems \cite{reips02}. For instance, the political voter mobilization experiment conducted by Robert et al. on Facebook was sharply criticized by a considerable number of scholars, who argued that scientific research should not interfere with national politics or users\textquotesingle voting behavior \cite{bond12}. Therefore, inappropriate design on actual large-scale social networks is likely to cause social injustice or infringe on the privacy of users, which violates the ethics of scientific research and may even be illegal. All these issues are due to improper intervention in the behavior of social network users and the collection of their private data.

In 2018, the Deep Mind team published a breakthrough study. By training the recurrent neural network, researchers simulated artificial agents with the function of autonomous mobility and judgment of spatial location in two-dimensional virtual space \cite{banino18}. These artificial agents produced a structure similar to that of the neural network cells of humans. The study provides a theoretical basis for using artificial intelligence technology to simulate human social behavior. If artificial agents can be used to replace real users to conduct experiments on large-scale social networks, potential moral and legal risks can be avoided, and the cost and threshold of the experiments can be reduced.

\section{PROPOSED METHOD}

\subsection{Social bot design}

\begin{figure*}
\includegraphics{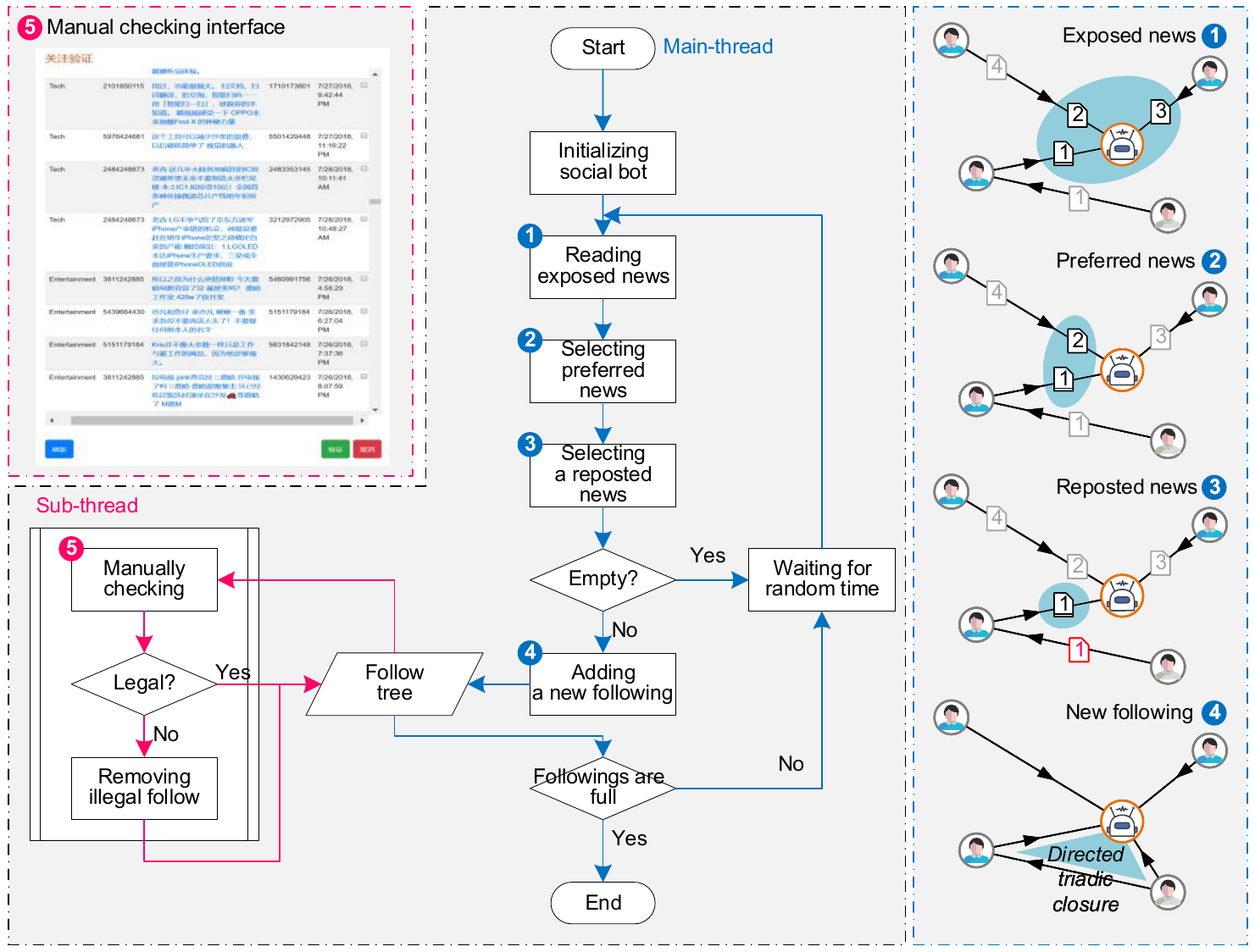}
\caption{Design of social bot.}
\label{fig:bot}
\end{figure*}

A social bot is a computer algorithm that automatically interacts with humans on social networks, trying to emulate and possibly alter their behavior \cite{ferrara16}. The social bot in our experiment is designed to imitate similarity-based relationship formation, which reflects the human behavior of selecting information and relationships depending on one\textquotesingle s preferences (i.e., self-selection).

The social bot is based on two well-known sociological hypotheses. The first is triadic closure \cite{bianconi14}. Triadic closure suggests that among three people, A, B, and C, if a link exists between A and B, and A and C, then there is a high probability of a link between B and C. The hypothesis reflects the tendency of a friend of a friend to become a friend and is a useful simplification of reality that can be used to understand and predict the evolution of social networks. Because the relationships embedded in Weibo should be modelled by a directed graph, the bots use directed triadic closure to expand their followings (Figure~\ref{fig:bot}) \cite{romero10}. The second hypothesis is homophily \cite{dandekar13}, i.e., people who are similar have a higher chance of becoming connected. In online social networks, for example, an individual creating her new homepage tends to link it to sites that are close to her interests (i.e., preferences).

Based on the two hypotheses, the workflow of the bots includes five steps (Figure~\ref{fig:bot}). (1) Initially, each bot is assigned 2 or 3 default followings, who mostly post or repost messages consistent with the preference of the bot. (2) A bot will periodically awaken from idleness at a uniformly random time interval. When the bot awakened, it can view the latest messages posted or reposted by its followings. As well known, not all unreaded information can be exposed to users of social networks. A bot can assess only the latest 50 messages (i.e., the maximum amount in the first page on Weibo) after waking up. Please note that we exclude the influence of algorithmic ranking and recommendation systems (i.e., pre-selection) on the information exposure by re-ranking all possible messages according to the descending order of posting time. (3) After viewing the exposed messages, the bot selects only the messages consistent with its preferences. The step depends on the FastText text classification algorithm, and all classification results from the algorithm are further verified by the experimenters to ensure correctness. (4) If there are reposted messages in selected messages, according to directed triadic closure, the bot randomly selects a reposted message and follows the direct source of the reposted message. Please note that Weibo limits direct access to the information about the followings and followers of a user; thus, bots must find new followings through reposting behavior. (5) If the following number reaches the upper limit, the bot stops running; otherwise, the bot becomes idle and waits to wake up again. To avoid legal and moral hazards, the bots in the experiment do not produce or repost any information.

Technically, the bot is built using the Selenium WebDriver API (http://www.seleniumhq.org/projects/webdriver) to drive Google Chrome Browser as a human to navigate the Weibo website. To train the FastText classifier for identifying preferred contents \cite{joulin16}, we use the THUCNews dataset (http://thuctc.thunlp.org), which contains approximately 740 thousand Chinese news texts from 2005 to 2011 and has been marked for 14 classes. We further combine and filter the original classes in the dataset to meet our requirement for recognizing Weibo text. All codes and data for building bots are freely accessible in GitHub (anonymous for review).
 
\subsection{Experiment design}

We are interested in exploring the following overarching question: what role do the individual preferences of social media users play in their news consumption? Based on the above intelligent social bots, we designed a controlled experiment to help us observe the news consumption and personal social network evolution of the bots with a specific preference. For each bot, we mainly measured the following response variables:

V1: The probability that the bot is exposed to the preferred content. For a given time $t$, the probability can be quantified by the preferred content ratio (PCR), which is the ratio of the amount of information matching its preference to the total amount of information from $0$ to $t$. Since the running of each bot is not completely synchronized, we normalize the time scale from $0$ to $1$, where $t=0$ represents the initial state after given the initial user, and $t=1$ represents the time when the bot ends the running.

V2: The distribution of word frequency in the exposed content. For a word, the word frequency is defined as the ratio of the number of messages that contain the word to all exposed messages of a bot.

V3: The attributes of all followings of a bot. These attributes include the proportion of followings with the same preference (the preference of followers is judged according to their screen name as well as the tags and content of their microblogs) and the number of the followings\textquotesingle followers and followings.

V4: The structure of the followings network of a bot (i.e., the personal social network of the bot), including clustering coefficient, in- and out-degree distribution, in- and out-degree correlation, and reciprocal ratio ($r$) \cite{boccaletti06,fagiolo07}. For a directed arc $<v_i,v_j>$, if there is a reversed arc $<v_j,v_i>$, is a reciprocal edge $(v_i,v_j)$ is defined between two nodes. As a result, the reciprocal ratio can be defined as:
\begin{equation}
  r=\sum^n_{i=1}\frac{n_e(v_i)}{n_a(v_i)}
\end{equation}
where $n_e$ is the number of edges linking to node $i$, and $n_a$ is the total number of arcs linking to $i$. We can also measure the correlation (called reciprocal correlation) between $n_e/n_a$ and $n_a$ to reflect the structural features of personal social networks.

For V1 and V2, we can quantify the extent to which the preference for particular kinds of content affects the diversity of new consumption. For V3 and V4, we can reveal the impact of users' preference on their personal social network structure and explore the relationship between the social network structure and content consumption.

We designed two experimental treatments and evaluated the difference in the above response variables. The two treatments were designed to reflect the variance in content preference on the Weibo. The two treatments are an entertainment group (EG) and a science and technology (sci-tech) group (STG). Bots in EG prefer entertainment messages, including celebrity gossip, fashion, movies, TV shows, music, and such, and bots in STG prefer sci-tech news, including nature, science, engineering, technological advance, digital products, Internet, and so on. Each treatment contained 34 bots who operated on the Weibo platform between 13 March and 28 September 2018. The initial followings come from a large enough user pool, which contained at least 100 candidates for each preference. The idle period of all bots was 2~4 hours. The maximum number of followings for each bot was 120, according to the Dunbar number.

Compared with the conventional method, our design is characterized by using intelligent bots as the agents to conduct experiments in a real online social environment and indirectly observe real human behavior. These bots act entirely on predefined behavior hypotheses and intelligent algorithms, needing manual operation by the experimenters only in the judgment of preference classification. Therefore, our experiment can be regarded as a type of randomized assignment.

\subsection{Standard for preference classification}

The most significant feature of the social bots is that they can autonomously identify whether information matches their preferences, benefiting from automatic and manual text classification. Therefore, the criteria of text classification are fundamental. We defined a content classification standard for the two kinds of content in our experiment (Table~\ref{tab:standard}).

\begin{table}
  \caption{Standard for text classification}
  \label{tab:standard}
  \begin{tabular}{p{1.3cm} p{3cm} p{3cm}}
    \toprule
    Standard & Entertainment & Sci-Tech\\
    \midrule
    Accept & (1) celebrity gossip, fashion, movies, TV shows, and pop music; (2) Explicitly containing the name, account or abbreviation of entertainers. & (1) nature, science, engineering, technological advances, digital products, and Internet; (2) technical company and university. \\
    \midrule
    Common reject & \multicolumn{2}{p{6cm}}{(1) commercial advertisement; (2) less than 5 Chinese characters.} \\
    \midrule
    Specific reject & (1) ACG content (i.e., animation, comic, and digital game); (2) art, literature, and personal feeling; (3) simple lyrics and lines; (4) personal leisure activities. & (1) financial or business report of technical or Internet company; (2) price of digital products; (3) military equipment; (4) daily skills; (5) constellations and divination; (6) weather forecast; (7) environmental conservation; (8) documentary with irrelevant content. \\
    \bottomrule
  \end{tabular}
\end{table}

The classification criteria shown in Table~\ref{tab:standard} were used to select initial followings, prepare the training corpus of the FastText classifier, manually verify, and classify exposed texts. The unified standard can ensure consistency in the judgment of preference and the classification of text.

\section{RESULTS}

\subsection{User preference drives polarization of content}

In social media, users\textquotesingle self-selection behavior of information source and information has been considered the primary mechanism of polarization. In this experiment, social bots were designed to expand personal social networks based on the two hypotheses, which aimed to simulate users\textquotesingle self-selection. We aimed to validate whether the effect of self-selection is identical under difference preferences.

\begin{figure}
\includegraphics{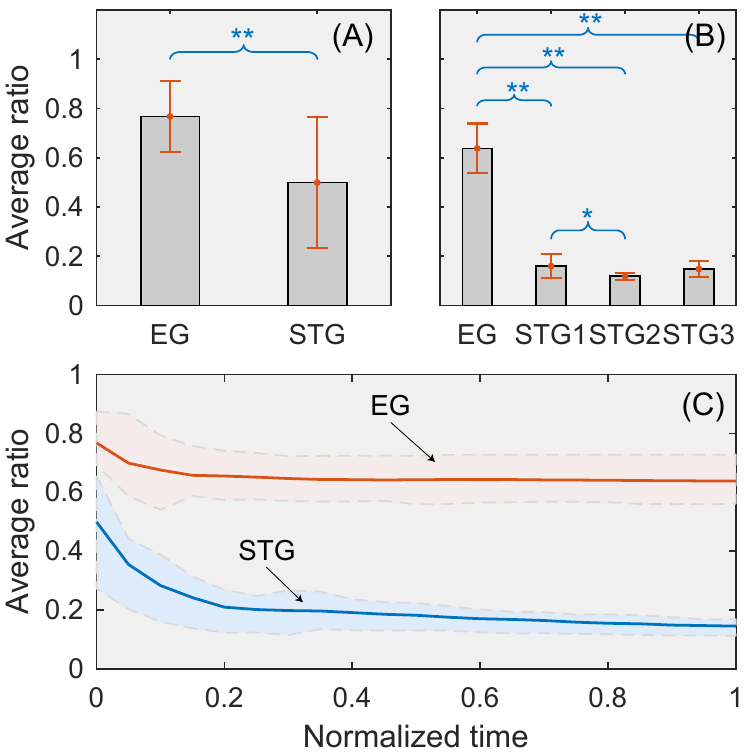}
\caption{Preference drives the polarization of exposed content. (A) In the initial state, there is an approximately 50\% higher preferred content ratio (PCR) in the entertainment (EG) than in the sci-tech (STG) treatment ($**p<0.01$, two-sided Mann-Whitney U test, $n=34$ bots). (B) In the final state, the PCR of the EG is 4.4 times that of the STG treatment ($**p<0.01$, two-sided Mann-Whitney U Test, $n=34$ bots). Furthermore, we divided bots in STG into three subgroups according to their initial PCR: STG1 ($0.6\leq PCR$, $n_1=14$), STG2 ($0.3\leq PCR<0.6$, $n_2=11$), and STG3 ($PCR<0.3$, $n_3=9$). Only STG1 and STG2 display a significant difference ($*p<0.05$, Kruskal-Wallis with Mann-Whitney post hoc test). (C) PCR changes with time. The filling areas indicate the range between the 25th and 75th percentiles of the ratios in the two treatments.}
\label{fig:pcr}
\end{figure}

For all 68 social bots, since the initial users have been set as the user who mainly reposts the corresponding preferred content, the PCR of EG and STG in the initial state can be as high as 76.77\% and 50.0\% on average (Figure~\ref{fig:pcr}(A)). That is, most of the messages exposed to both treatments are concentrated in the corresponding preference. Even so, preference has a significant impact on information diversity. The PCR of EG was significantly higher than that of STG ($p<0.01$, two-sided Mann-Whitney U test). This initial difference between EG and STG provides a baseline for our further analysis.

Compared to the initial state, we are more concerned about the diversity of messages consumed by social bots after personal social networks have formed (Figure~\ref{fig:pcr}(B)). As the PCR of STG in the initial state is significantly less than that of EG and has a large variety (Figure 3A), we further divide STG into three subgroups according to their initial PCR to explain the effect of initial deviations: STG1 (initial PCR is more than 0.6, 14 bots), STG2 (initial PCR is between 0.3 and 0.6, 11 bots), and STG3 (initial PCR is less than 0.3, 9 bots). The results show that when the personal social network reaches 100 to 120 followings, EG still has the higher PCR with an average of 63.78\%, while the average PCR of the three STG subgroups is lower than 20\%. Moreover, Only STG1 and STG2 display a significant difference ($*p<0.05$, Kruskal-Wallis with Mann-Whitney post hoc test). The results suggest that regardless of the initial state, bots in the STG treatment can receive diverse information rather than sci-tech content. From the changes of PCR with the growth of personal social networks, we can also observe the effect of user preference on information polarization (Figure~\ref{fig:pcr}(C)). With the social network grows, the PCR of EG declines only slightly in the early stage and remains unchanged after the further increase of followings. In STG, new followings can continuously reduce the PCR; that is, they can bring more nonpreferred content. Therefore, there is a correlation between the users' preferences and the polarization of their news consumption in Weibo.

\begin{figure*}
\includegraphics{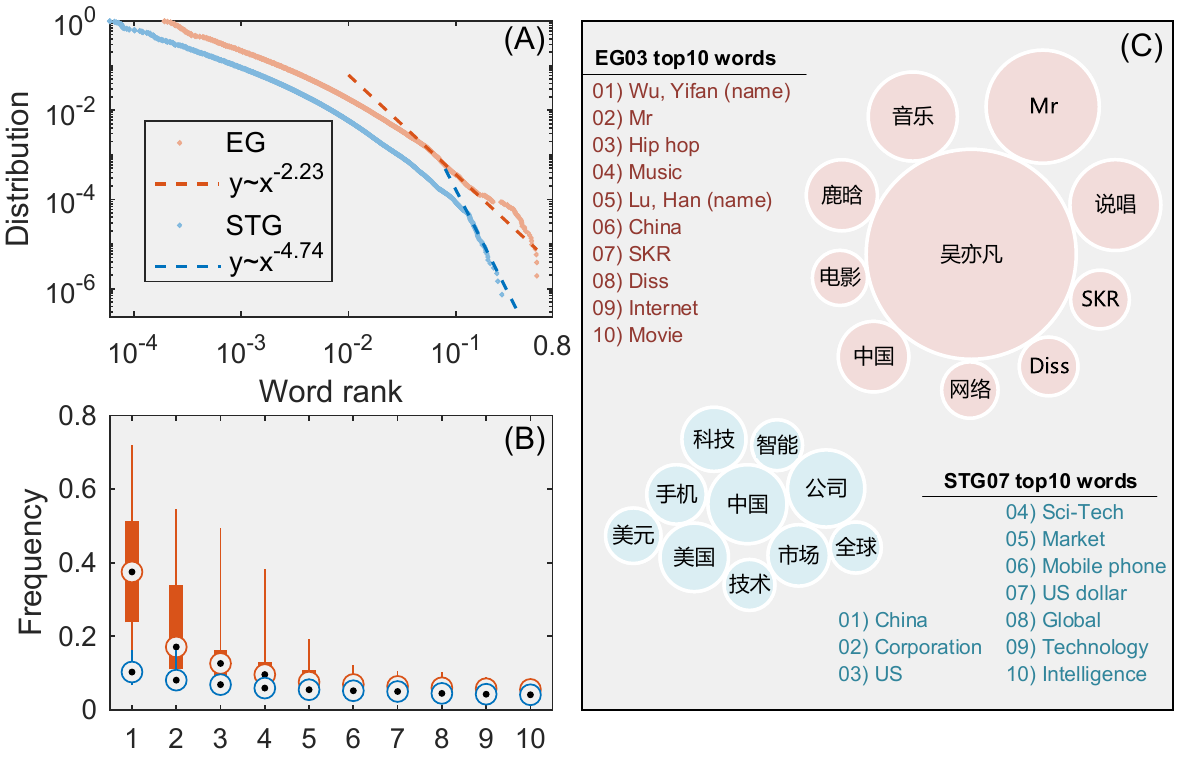}
\caption{Semantic analysis of exposed contents. (A) Inverse cumulative distribution of word frequency (log-log plot). The EG treatment has a longer tail than the STG treatment; i.e., the portion of the EG distribution has more higher-frequency words. (B) Box plots of the frequency of top 10 words in both the EG and STG treatments. (C) Demonstration of top 10 words in exposing messages from EG 03 bot and STG 07 bot. The size of the bubble indicates the word frequency.}
\label{fig:word}
\end{figure*}

The effect of preference on content consumption is reflected not only on the type of content but also on the semantic level. In both EG and STG treatments, the word frequency of the preferred content exposed to the social bots presents a power-law distribution (Figure~\ref{fig:word}(A)). The power-law distribution ($y\sim x^{-\alpha}$) indicates that there are some dominant words with high word frequency in the preferred content. Moreover, the dominance of a few words is more severe in EG than in STG ($\alpha=2.23$ for EG and $\alpha=4.74$ for STG). In the EG treatment, the maximum frequency of a word can be up to 0.4 on average, which means that the same word can be found in 40\% of messages on average. However, in the STG treatment, the maximum word frequency is no more than 0.15 (Figure~\ref{fig:word}(B)). Even among the top 10 words of EG, the first and second words are more dominant than other words. By analyzing the top 10 high-frequency words of the two treatments, we find that the high-frequency words of EG are usually the names of certain entertainment celebrities and related specific words, while the high-frequency words of STG are some common words, such as China, America, technology, market, and company (Figure~\ref{fig:word}(C)). From this, it can be seen that the messages received by bots preferring entertainment content not only concentrate on a single type but also show a univocal trend in semantics.

\subsection{The personal social network}

The dissemination mechanism of Weibo makes the followings of a bot become the primary and direct information sources for the bot. Therefore, the personal social network of a social bot (i.e., the network consisting of its followings) is its media environment and reflects a local structure to diffuse specific types of information. We observed the personal social network of each bot and found that the bots present different structures according to their preferences.

\subsubsection{The attributes of followings}

\begin{figure}
\includegraphics{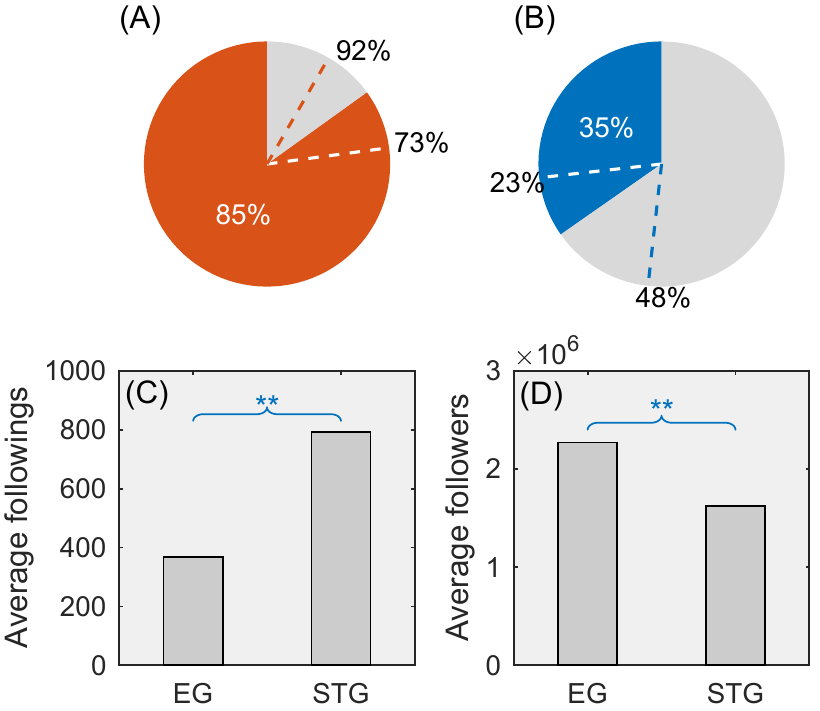}
\caption{Demographic features of the followings of bots. (A) and (B) Percentage of followings with the same preference with bots in the EG and STG treatments. The white and colored dotted lines indicate the minimum and maximum percentages, respectively. (C) The average number of followings of followings ($**p<0.01$, two-sided Mann-Whitney U Test, 2,675 followings in EG and 1,622 followings in STG). (D) The average number of followers of followings ($**p<0.01$, two-sided Mann-Whitney U Test, 2,675 followings in EG and 1,622 followings in STG). Due to the high variance, we do not show error lines in this plot.}
\label{fig:user}
\end{figure}

The followings of a bot are the nodes of its personal social network. Our results showed that in the EG treatment, the average proportion of nodes with the entertainment preference is 85.32\% (up to 92.14\% and down to 73.47\%) (Figure~\ref{fig:user}(A)). The high proportion suggests that, with the entertainment preference, users\textquotesingle self-selection can effectively filter the information source, making most of the information sources consistent with their preference and thus resulting in a filter bubble. In the STG treatment, however, the proportion is only 35.33\% on average (up to 47.86\%, down to 23.12\%) (Figure~\ref{fig:user}(B)). Based on the information diffusion mechanism of Weibo and the design of social bots, a low proportion indicates that users with the sci-tech preference have a high possibility of following users with other preferences, and users with various preferences could forward the sci-tech content. The interaction between multiple preferences effectively enhances the diversity of news consumption. For the sci-tech preference, as a result, the content-based self-selection behavior is inefficient in developing information polarization.

Not only the composition of followings but also the features of followings in the two treatments are significantly different. Figure~\ref{fig:user}(C) and (D) show that, compared with STG, EG has 53.78\% fewer followings (i.e., an indirect information source for bots) but 23.43\% more followers on average ($p<0.01$, two-sided Mann-Whitney U Test). In other words, users with entertainment preference tend to interact with fewer news sources while diffusing information to a large number of audiences. The ``trumpet-like'' feature can incite information polarization. Therefore, users' preferences not only directly affect news consumption through self-selection but also change their social networks to reinforce this effect.

\subsubsection{The connection of personal social networks}

\begin{figure}
\includegraphics{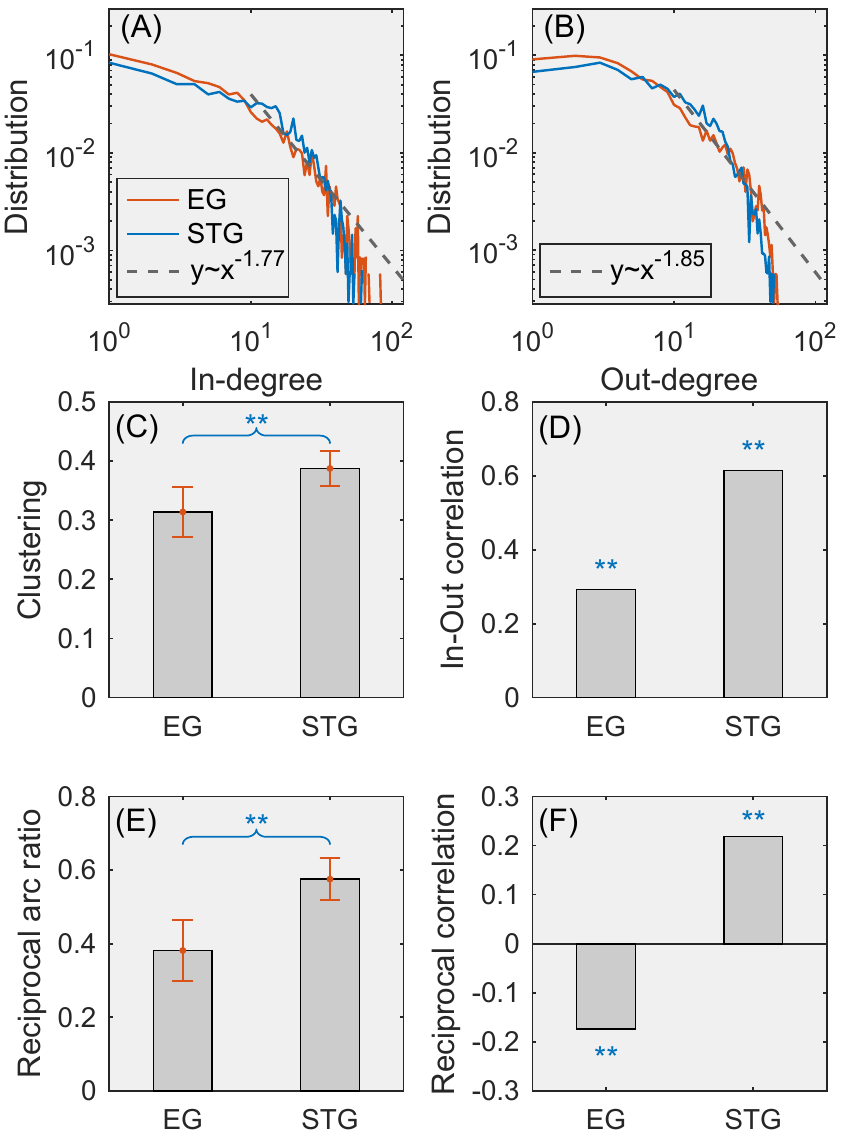}
\caption{Structural features of personal social networks. (A) Mean of inverse cumulative distribution of in-degree (log-log plot). (B) Mean of inverse cumulative distribution of out-degree (log-log plot). (C) Personal social networks in STG have a higher clustering coefficient than networks in EG ($**p<0.01$, two-sided Mann-Whitney U Test, $n=34$ networks). (D) Compared with EG, nodes in STG present a stronger correlation between in- and out-degrees ($**p<0.01$, Pearson correlation coefficient, 54,823 nodes for EG, 51,838 nodes for STG). (E) Personal social networks in STG obtain more reciprocal edges than networks in EG ($**p<0.01$, two-sided Mann-Whitney U Test, $n=34$ networks). (F) Nodes in EG show a weak negative correlation between reciprocal edge degree ($>0$) and total degree. In contrast, nodes in STG show a positive reciprocal degree correlation ($**p <0.01$, Pearson correlation coefficient, 3,546 nodes for EG, 3,368 nodes for STG).}
\label{fig:net}
\end{figure}

\begin{figure*}
\includegraphics{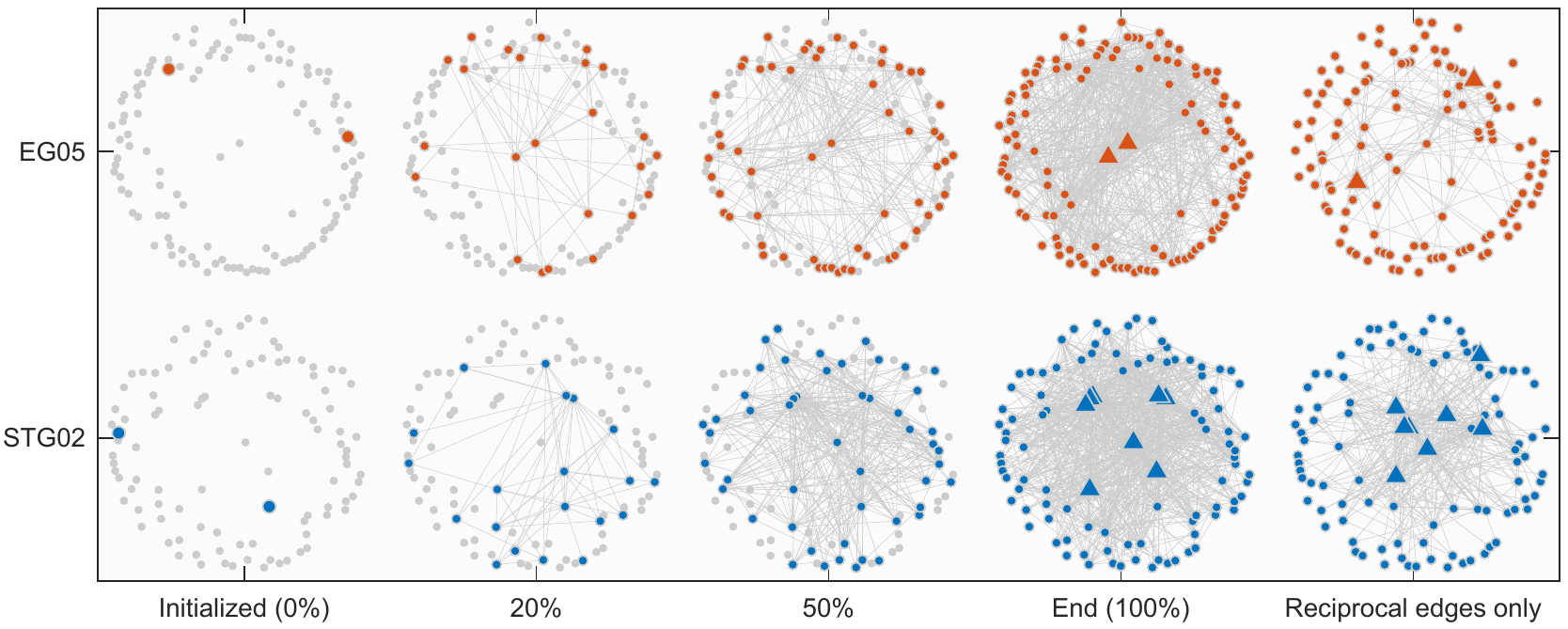}
\caption{ Visualization of the evolution of personal social networks. In this demonstration, Social bot 05 in EG (EG05) and bot 02 in STG (STG02) display a similar growth process from initialized two followings to the end of running. However, STG02 obtains more reciprocal edges than EG05. The triangles represent the nodes with high in-degree, that is, the corresponding users have a large number of followers. The visualization is based on the radial layout with in-degree centrality; thus the node with higher in-degree is closer to the center of the plot.}
\label{fig:visual}
\end{figure*}

In addition to the followings' attributes, the structure of the personal social network is also modified by the user preference. By roughly comparing the in- and out-degree distributions of EG and STG\textquotesingle s personal social networks, we find that both present power-law distributions ($\alpha=1.766$ for in-degrees and $\alpha=1.847$ for out-degrees) and are almost identical (Figure~\ref{fig:net}(A) and (B)). Furthermore, the visualization of network evolution also shows that EG and STG have similar personal social networks (Figure~\ref{fig:visual}). However, additional structural features reveal the opposite result.

Figure~\ref{fig:net}(C) shows that the average clustering coefficient of STG networks is 35\% higher than that of EG networks ($p<0.01$, two-sided Mann-Whitney U Test). Given the low level of information polarization in STG, this is an unexpected result, as a dense community spreads diverse information. This counterintuitive phenomenon can be attributed to differences in the connection pattern of personal social networks. Figure~\ref{fig:net}(D) shows that there is a high positive correlation (0.6) between the in- and out-degrees of STG networks, while the degree correlation of EG networks is very weak (0.2). This correlation implies a difference between the two types of networks on reciprocal edges. We found that an average of 60\% of the arcs in STG networks belong to the reciprocal edge, while the proportion is only 40\% in EG (Figure~\ref{fig:net}(E)). Besides, in STG, the degree of reciprocal edge presented a weak positive correlation (0.25) with nodes' total degree; that is, the higher-degree nodes might have more reciprocal edges (Figure~\ref{fig:net}(F)). In EG, the degree of the reciprocal edge is negatively correlated with the total degree (-0.18); that is, the higher-degree nodes prefer the one-way relationships in Weibo (Figure~\ref{fig:net}(F)).

The visualization analysis also shows the same results. If all nonreciprocal arcs are removed, we find that the STG networks still have a high connection density, and the high-degree nodes are still maintained in the centrality, while EG networks are the opposite, with the sparse connection, and the high-degree nodes are marginalized (Figure~\ref{fig:visual}). Based on the above results, an EG network tends to be a unidirectional star-like structure with a few high-degree nodes, while an STG network tends to be a bidirectional clustering structure. The unidirectional star-like structure means selecting one or several central nodes to play the role of broadcasters to send information to all other nodes with few interactions between them. However, the diffusion path of this structure is mainly dependent on the selected central node. If the central nodes in the star-like structure produce only a narrow set of news, the others will be limited to a lower diversity of content. This selection effect of central nodes tends to enlarge the polarization of the content \cite{loreau01}. In the bidirectional clustering structure, all nodes are placed in a clustering group, and most connections are bidirectional. Those nodes can efficiently exchange information and realize a complementary effect with their friends with distinct preferences. The complementary effect can promote the diversity of information \cite{loreau01}.

Although all social bots select potential followings based on preferred content to ensure that these users post or repost the content they like, whether the self-selection behavior can narrow users' news consumption is dependent on the users' preferences. Moreover, different news consumption patterns are accompanied by different compositions and structures of personal social networks.

\section{DISCUSSION}

In this paper, we create a new controlled experiment approach to reveal the link between user preferences and patterns of news consumption on a Chinese microblogging platform (Weibo). This method can be seen as a real-time user behavioral simulation on real-world social networks using the computer and artificial intelligence technology. (1) The method regards real-world social networks as an objective media environment and records and analyzes the feedback of the environment to specific user behaviors. (2) This method uses artificial intelligence technologies (e.g., natural language processing) to simulate specific user behaviors instead of employing volunteers or collecting users' private data. The approach not only improves the freedom of the experimenters in designing and controlling the experiment but also avoids the ethical risk of violating users' privacy. (3) Compared with the experimental method of directly altering the real-world social network \cite{bond12}, this method gives the experimenter more flexibility and allows more researchers to conduct their own experiments without internal assistance from social network operators. (4) Compared with the experimental method based on artificial social networks \cite{centola10,centola11}, this method can work directly on real-world social networks and can control user behaviors more directly with lower cost. However, limited by the current technical conditions, this method can intelligently simulate only relatively simple user behaviors. How to simulate complex user behaviors (especially initiative feedback behaviors such as comments and chats) is still a challenging task.

Taking advantage of our experimental method, we compare the news consumption of users with different preferences. Previous studies on the polarization mechanism focus on the effect of users' self-selection behaviors \cite{kakiuchi18}. First, a large number of studies focus on the dissemination of different opinions in response to particular content, such as political news \cite{schmidt17,conover11,markus13,narayanan18}. In this case, users are actually in the same media environment, and the differences in content preferences can be ignored. Second, the rise of pre-selection technologies (e.g., recommendation systems) makes people more attentive to the comparison of pre-selection and self-selection than to the impact of the media environment \cite{bakshy15}. However, our research shows that user preferences might be the primary factor affecting the diversity of news consumption. Distinct preferences can lead to entirely different news consumption patterns even with the same self-selection behavior. Such results deepen our understanding of the mechanism of information polarization. Furthermore, in this experiment, we make the ideal assumption that each user has only a single preference. However, people usually have multiple preferences. How the interaction of multiple preferences shapes the diversity of news consumption remains an open question.

Our results also show that user preference not only affects information diversity but also shapes personal social networks. Depending on the mechanism of Weibo and the design of a social bot, the messages that a bot can receive is limited to the followings\textquotesingle posts and reposts. Therefore, information diversity depends on followings\textquotesingle self-selection and the structure of personal social networks. The impact of network structure on information diffusion is well known \cite{boccaletti06,min13,min14}, but people often overlook the role of content in it. Until now, only a few works have noticed the link between content and structure \cite{teng12,schmidt17}. In communication studies, the media ecology theory has revealed the connection between the form and structure of media and content \cite{scolari12}. They even think that ``the medium is the message.'' \cite{mcluhan67}. Our results show that even in the same media, different content should spread in distinct sub-structures. For example, we found that personal social networks of users with entertainment preferences tend to be a unidirectional star-like structure, which is more powerful for the broadcasting of narrow content. It can be seen that user preference drives the coevolution of content consumption and personal social network structure, and the coevolution strengthens information polarization.

Our work provides valuable insights into the theoretical analysis of diffusion dynamics. We found that, even in the same network, information about different contents should be propagated through different sub-structures. This phenomenon has led us to re-examine the results of previous studies based on a unified macroscopic structure, such as small-world networks or scale-free networks \cite{boccaletti06}. Fortunately, researchers have begun to use multilayer network models to describe the differences in sub-structures. However, the multilayer network model for content differences and the related propagation mechanisms remain to be studied. Furthermore, our work is also instructive for the design of the recommendation system \cite{zhou10}. Users with different preferences need different recommendation strategies. For example, our results show that users preferring entertainment content need more diverse recommendations to avoid polarization, while users preferring sci-tech content need more relevant content to maintain their activity in social networks \cite{min17}. As a result, the recommendation system not only needs to pay attention to individual preferences and behaviors but also needs to understand the collective behavior of the user groups with the same preferences.

Our work is a preliminary experimental analysis of the relationship between user preference and news consumption, so there are still many limitations. First, the impact of initial users requires further evaluation. In the experiment, it was difficult for us to find users who post merely sci-tech content ($RCP\geq0.8$) as the initial followings of bots. Although we find that the weak selection intensity does not affect the result, the effect of the initial values still requires a more detailed analysis. Second, limited by the accuracy of the text classification algorithm, the robot currently only selects potential followings through the received real-time messages. The selection is relatively loose. If the robot can examine all the information about potential followers (historical posts, tags, avatars, etc.), the effect of rigorous selection requires further study. Third, the two types of preferences are not enough to explain all the problems. The study of more preferences (e.g., sport, finance, politics, etc.) will help us to understand the complexity of the triad of preference, content, and structure in news consumption.

In sum, we have introduced a new experimental approach using social bots and revealed the collective behavior behind information polarization, that is, a group of users with the same preference can form a media environment facilitating polarization by specialized structure and selection behaviors.

\begin{acks}
This work was supported by the National Natural Science Foundation of China (Grant Nos. 71303217, 31370354 and 31270377) and the Zhejiang Provincial Natural Science Foundation of China (Grant No. LY17G030030, LGF18D010001, LGF18D010002).

We thank Professor Jie Chang and Professor Ying Ge (Zhejiang University) for their guidance on the diversity theory.

We thank Aizhu Liu, Chenyi Fang, Conger Yuan, Fan Li, Hao Li, Hao Wu, Haochen Hou, Hengji Wang, Jian Zhou, Jiaye Zhang, Jinmeng Wang, Junhao Xu, Lingjian Jin, Longzhong Lu, Lu Chen, Luchen Zhang, Qiuhai Zheng, Qiuya Ji, Renyuan Yao, Ruonan Zhang, Shang Gao, Shicong Han, Songyi Huang, Ting Xu, Wei Fang, Wei Zhang, Xingfan Zhang, Yijing Wang, Yingjie Feng, Yinting Chen, Yiqi Ning, Yujie Bao, Yuying Zhou, Zheyu Li, Ziyu Liu for their contribution to the manual text classification.

We thank Haidan Yang for her introduction to the theory of communication.
\end{acks}

\bibliographystyle{unsrt}
\bibliography{www}

\end{document}